\documentclass[aps,prl,twocolumn,superscriptaddress,floatfix]{revtex4-2}
\usepackage{bm}
\usepackage{amsmath}
\usepackage{amssymb}
\usepackage{amsfonts}
\usepackage{graphicx}
\usepackage{color}
\usepackage[T1]{fontenc}
\DeclareMathOperator{\Tr}{Tr}
\newcommand{\mirror}[1]{\overline{#1}}
\newcommand{\rt}{\tilde\rho}

\begin{document}

\title{Kramers' Theorem Leaves a Topological Footprint in Classical Records}

\author{Chon-Fai Kam}
\email{dubussygauss@gmail.com}
\affiliation{Dipartimento di Fisica e Chimica ``Emilio Segr\`e'', Universit\`a degli Studi di Palermo, Via Archirafi 36, I-90123 Palermo, Italy}
\affiliation{Universit\'e Paris Cit\'e and Universit\'e de La R\'eunion, 75015 Paris, France}

\author{Kai-Wen Wong}
\affiliation{Life Actuarial Department, Taiwan Insurance Institute, 6th floor, No.~3, Nan-Hai Road, Taipei 100, Taiwan (R.O.C.)}

\date{\today}

\begin{abstract}
Time reversal $\Theta$ squares to $-1$ on a half-integer spin, so every state is orthogonal to its time reverse. This sign has been read only from quantum objects. The equilibrium state loses it, but the classical record keeps it. For a detector that resolves handedness, the probabilities that one stretch of record is followed by the time reverse of another measure the cohomology class of time reversal, $\Theta^2=\pm1$. Their matrix carries this sign in its signature, a topological integer that survives equal emission and absorption losses. A programmed emitter on quantum processors reads it in about $10^3$ shots.
\end{abstract}

\maketitle

The antiunitary time-reversal operator $\Theta$ squares to $1$ on an integer spin and to $-1$ on a half-integer spin. The sign $-1$ makes the degeneracy of every level of a time-reversal invariant spin even, and a generic level a Kramers doublet \cite{kramers1930,wigner1932}. This sign has been read from level statistics \cite{dyson1962}, from weak antilocalization \cite{hikami1980} and from the bands of topological insulators \cite{kanemele2005}. It has also been read from the entanglement and the partial inversion of quantum states, including states reconstructed by tomography or randomized measurements \cite{pollmann2010,pollmann2012,choo2018,elben2020}. All these readings use quantum objects, and none uses the classical record of clicks that a monitored system writes by itself.

A monitored spin relaxes to equilibrium, where each Kramers doublet is unpolarized and looks the same for either sign of $\Theta^2$. The sign multiplies state vectors and drops out of density matrices, so the record of such a spin seems unable to carry it. The detector clicks, however, carry time correlations that the equilibrium state does not hold. Microscopic reversibility relations connect the probabilities of a record and its time reverse \cite{onsager1931,casimir1945,crooks2008,horowitz2013,manzano2018}. These relations and the quantum detailed-balance conditions behind them either fix $\Theta^2=1$ or never use its sign, since the ratio $P(x)/P(\mirror{x})$ never needs it \cite{fagnola2010,manzano2015,benoist2025}. Symmetry classes of monitored dynamics are defined by symmetries of the individual Kraus operators and leave physical time reversal aside \cite{xiao2026}.

\begin{figure*}[t]
\includegraphics[width=\textwidth]{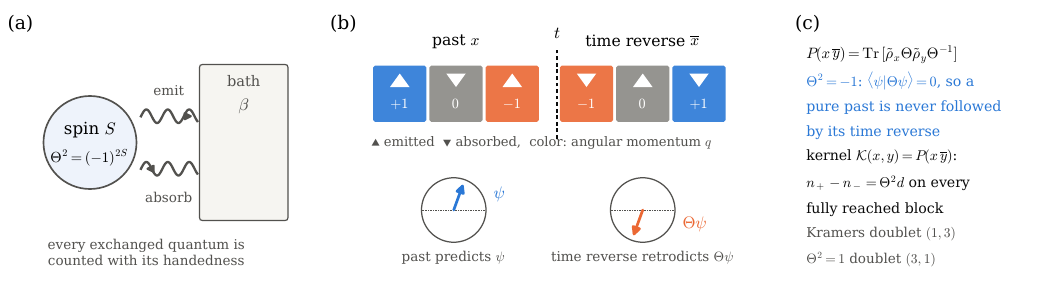}
\caption{(a) An equilibrium spin exchanges quanta with a bath at inverse temperature $\beta$. (b) A record and its time reverse across the time $t$. The time reverse reads the past backwards with emission and absorption exchanged. The past predicts a state $\psi$, and its time reverse retrodicts $\Theta\psi$. (c) The probability of the pair is the overlap of $\psi$ and $\Theta\psi$ [Eq.~\eqref{eq:mirror}] and vanishes for $\Theta^2=-1$. The matrix of such probabilities has signature $\Theta^2d$ [Eq.~\eqref{eq:signature}].}
\label{fig:concept}
\end{figure*}

Here we show that the sign appears in the probability that a stretch of record is followed by its time reverse, the same stretch read backwards with emission and absorption exchanged (Fig.~\ref{fig:concept}). This probability equals the overlap of the state the stretch prepares with its own time-reversed image. It gives a selection rule, since for $\Theta^2=-1$ a stretch that prepares a pure state is never followed by its time reverse. A half-integer spin in thermal equilibrium realizes the rule when a detector records the handedness of each emitted or absorbed quantum. A detector of linear polarization unravels the same master equation but cannot tell the two Kramers partners apart, so the sign in the record depends on the detector as well as on the emitter. The probabilities that one stretch is followed by the time reverse of another form a matrix whose signature is fixed by $\Theta^2$ and survives losses that treat emission and absorption alike, so the sign is a property of the data. A classically microreversible model of the same record needs a pair of microstates for each negative eigenvalue, so a Kramers doublet costs six classical states against two quantum ones. The record thus measures the cohomology class of time reversal, an element of $H^2(\mathbb{Z}_2^T,U(1))=\mathbb{Z}_2$ for the group $\mathbb{Z}_2^T$ generated by $\Theta$ \cite{chen2013}. A programmed emitter on current quantum processors can read it with about $10^3$ shots.

\textit{Setting.}---A detector registers the outcomes $m$ of an emitter, a quantum system that exchanges quanta with its surroundings. Each outcome updates the emitter's state through a Kraus operator $B_m$. A stretch of record $x=m_1m_2\cdots m_n$ has the Kraus operator $M_x=B_{m_n}\cdots B_{m_1}$. The emitter starts in its unique stationary state $\rho_J$ and produces $x$ with probability $P(x)$. Time reversal maps each outcome $m$ to an outcome $\tau(m)$ with $\tau(\tau(m))=m$ and turns emissions into absorptions. We call $\mirror{x}=\tau(m_n)\cdots\tau(m_1)$ the time reverse of $x$. The record is reversal-symmetric if $P(x)=P(\mirror{x})$ for every stretch $x$. This holds when the reversed operations $\tilde B_m=\rho_J^{1/2}B_m^\dagger\rho_J^{-1/2}$, which emit each stretch backwards with its original probability \cite{crooks2008}, obey
\begin{equation}
\tilde B_m = e^{i\varphi_m}\,\Theta\,B_{\tau(m)}\,\Theta^{-1},\; \Theta=UK,
\label{eq:relation}
\end{equation}
where the phase $\varphi_m$ is arbitrary, $U$ is unitary and $K$ is complex conjugation \cite{pra}. This relation is quantum detailed balance with time reversal \cite{fagnola2010,manzano2018}. Conversely, a reversal-symmetric record implies Eq.~\eqref{eq:relation} when the dynamics is irreducible and the record informationally complete \cite{benoist2025}. Comparing the counts of each stretch with those of its time reverse thus tests the relation.

\begin{figure}[tbp]
\includegraphics[width=\columnwidth]{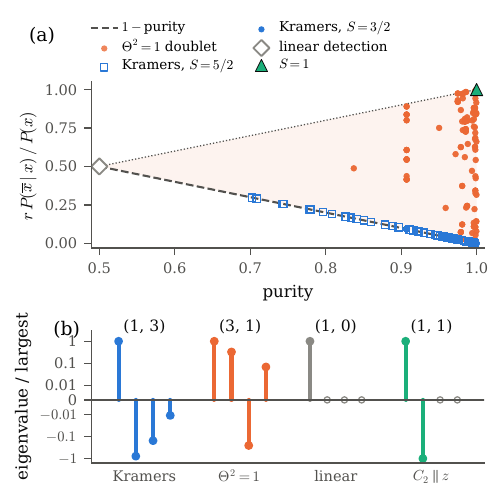}
\caption{(a) Rescaled probability $rP(\mirror{x}|x)/P(x)$ that a stretch $x$ is followed by its time reverse, against the purity $\Tr\hat\rho_x^2$ of the state that $x$ prepares, for stretches of up to four jumps (three for $S=5/2$). Kramers emitters with $S=3/2$ and $5/2$ fall on the line $1-\Tr\hat\rho_x^2$ of Eq.~\eqref{eq:doublet}. The programmed doublet with $\Theta^2=1$ lies between this line and the dotted line of slope one, and a non-degenerate level sits at 1. Under linear detection the Kramers pseudospin stays maximally mixed. (b) The four largest eigenvalues of one kernel block divided by the largest. The pairs above each group give the inertia $(n_+,n_-)$. A tilted spin $3/2$ gives $(1,3)$ under circular detection and $(1,0)$ under linear detection. The $\Theta^2=1$ doublet gives $(3,1)$, and a spin $3/2$ with a twofold axis along $z$ gives $(1,1)$.}
\label{fig:rule}
\end{figure}

\textit{A selection rule.}---After a stretch $x$ the emitter is left in $\rho_x=M_x\rho_JM_x^\dagger$, whose trace is $P(x)$, and its normalized state is $\hat\rho_x=\rho_x/P(x)$. By Eq.~\eqref{eq:relation}, the time reverse of a stretch $y$ read with the stationary state as prior retrodicts $\Theta\hat\rho_y\Theta^{-1}$. The probability that $x$ is followed by the time reverse of $y$ is the overlap of the predicted and the retrodicted state (End Matter~\ref{em:proofs}):
\begin{equation}
P(x\,\mirror{y}) = \Tr\big[\rt_x\,\Theta\,\rt_y\,\Theta^{-1}\big],\; \rt_x\equiv\rho_J^{-1/4}\rho_x\,\rho_J^{-1/4}.
\label{eq:mirror}
\end{equation}
When $\Theta^2=-1$ and a stretch $x$ leaves the emitter in a pure state, $\rt_x$ has rank one and is orthogonal to its time reverse, so $x$ is never followed by its own time reverse (End Matter~\ref{em:proofs}). A jump through a rank-one operator leaves the emitter in a pure state whatever came before. Emitters with an axial splitting have such jumps \cite{pra}, and in other emitters a longer stretch leaves a purer state. On a Kramers doublet the stationary state is uniform, $\rho_J=r\openone$ with $r$ the weight of each state. For the doublet viewed as a pseudospin $\tfrac12$, time reversal reverses its direction, $\Theta\hat\rho_x\Theta^{-1}=\openone-\hat\rho_x$. The rule then becomes a law for the conditional probability,
\begin{equation}
P(\mirror{x}\,|\,x) = \frac{P(x)}{r}\big(1-\Tr\hat\rho_x^{\,2}\big).
\label{eq:doublet}
\end{equation}
This probability is proportional to the linear entropy of the state, the uncertainty that the record leaves within the Kramers pair. Since the linear entropy of a doublet never exceeds $\tfrac12$, a Kramers doublet obeys $P(\mirror{x}|x)\le P(x)/(2r)$. With energies resolved, the counts give $r$ as half the total probability of the stretches that end in the doublet, so a violation of this bound excludes $\Theta^2=-1$ from the counts alone. For a non-degenerate level of an integer spin the state is its own time reverse, and $P(\mirror{x}|x)$ equals $P(x)/r$. A doublet with $\Theta^2=1$ lies in between, $1-\Tr\hat\rho_x^2\le rP(\mirror{x}|x)/P(x)\le\Tr\hat\rho_x^2$ \cite{pra}. Figure~\ref{fig:rule}(a) shows the three cases.

\textit{A physical emitter.}---Equation~\eqref{eq:relation} holds for a spin $S\ge1$ whose levels are split by a time-reversal invariant Hamiltonian $H$, such as the zero-field splitting $H=DS_z^2+E(S_x^2-S_y^2)$ with axial and rhombic constants $D$ and $E$. The spin exchanges quanta with a thermal bath through its angular momentum, slowly enough for the secular approximation \cite{davies1974,breuer2002}. The jump operators $L_q(\omega)=\sum_{\varepsilon'-\varepsilon=\omega}\Pi_\varepsilon S_q\Pi_{\varepsilon'}$ lower the energy of the spin by $\omega$, where $S_q$ are the spherical components of the spin and $\Pi_\varepsilon$ is the projector onto the eigenspace of $H$ with energy $\varepsilon$. Their rates $\gamma(\omega)$ obey detailed balance at inverse temperature $\beta$, $\gamma(-\omega)=e^{-\beta\omega}\gamma(\omega)$. A detector records for each exchanged quantum the energy $\omega$ that the spin releases and the angular momentum $q\in\{0,\pm1\}$ that the quantum transfers to the spin along the $z$ axis. By time-reversal symmetry every state of a Kramers doublet decays at the same total rate, so averaging over the waiting time leaves a single Kraus operator for each outcome (End Matter~\ref{em:emitter}).

Read backwards, the emission of a quantum $\omega$ through $S_q$ becomes the absorption of the same quantum through $S_q^\dagger$, and detailed balance gives it the rate of a true absorption. Time reversal maps $S_q^\dagger$ to $-S_q$ and leaves the energies unchanged, so $\tilde B_{q,\omega}=-\Theta B_{q,-\omega}\Theta^{-1}$. This is Eq.~\eqref{eq:relation} with $\tau(q,\omega)=(q,-\omega)$. The $\Theta$ in this relation is the time reversal of the spin, so a half-integer spin gives a record with $\Theta^2=-1$. The derivation uses no conserved angular momentum, so Eq.~\eqref{eq:relation} holds for any tilt of the splitting axes from the $z$ axis \cite{pra}.

A detector of linear polarization cannot tell the two Kramers partners apart, so the pseudospin stays maximally mixed. Both detectors count the same quanta in different polarization bases, so they unravel the same master equation of the spin. Quantum trajectories and smoothed states depend continuously on the unraveling \cite{wiseman2010,laverick2025}. Switching from circular to linear detection nevertheless makes the reading of Eq.~\eqref{eq:signature} below jump from $(1,3)$ to $(1,0)$ [Fig.~\ref{fig:rule}(b)].

Reversal symmetry of the record requires an emitter that is time-reversal invariant and in equilibrium. A magnetic field violates the first condition, and once the detector resolves the Zeeman splitting a field of 1\% of the zero-field splitting changes $P(\mirror{x})/P(x)$ by order one. Baths at different temperatures violate the second, as they do for the Kramers degeneracy of open systems \cite{lieu2022}. If they still treat $q=1$ and $q=-1$ alike, each reversed jump stays the time reverse of its partner up to a numerical factor, so a stretch that prepares a pure state is still never followed by its own time reverse \cite{pra}.

\textit{A signature.}---The data alone do not fix the sign of $\Theta^2$ in Eq.~\eqref{eq:relation}, since a hidden spin $\tfrac12$ can flip it. Let each outcome $m$ of an emitter with $\Theta^2=1$ also rotate this hidden spin by $C_m\in SU(2)$, with $C_{\tau(m)}=C_m^{-1}$. The rotations are unitary and never change a probability, so the enlarged model produces the same record. Its time reversal includes that of the hidden spin, so the enlarged model has $\Theta^2=-1$ \cite{pra}. The data instead fix the kernel
\begin{equation}
\mathcal K(x,y)\equiv P(x\,\mirror{y})
\label{eq:kernel}
\end{equation}
on all stretches $x,y$ of a fixed length $n$. This is the Hankel matrix of the process with its future index time-reversed. For long enough stretches its rank is the smallest dimension of a linear model that generates the process \cite{monras2016}. The normalized kernel $\mathcal K(x,y)/\sqrt{P(x)P(y)}$ has eigenvalues whose moduli are the canonical correlations between past and future \cite{lindquist2015}. Their signs are the time-reversal parities that Casimir attached to Onsager's variables \cite{casimir1945} (End Matter~\ref{em:proofs}). By Eq.~\eqref{eq:mirror}, $\mathcal K$ is the Gram matrix of the real symmetric form $b(X,Y)=\Tr[X\,\Theta Y\Theta^{-1}]$ evaluated on the operators $\rt_x$. Stretches that leave the emitter in the same $\Theta$-invariant subspace form a block. On the Hermitian matrices of a block of dimension $d$, $X\mapsto\Theta X\Theta^{-1}$ is an involution with trace $\Tr(UU^*)=\Theta^2d$, the number of independent time-even operators minus the number of time-odd ones, and this is the signature of $b$ on the block. By Sylvester's law of inertia \cite{sylvester1852,horn2013}, on every block of dimension $d$ whose Hermitian operators the $\rt_x$ span,
\begin{equation}
\operatorname{rank}\mathcal K=d^2,\; n_+-n_-=\Theta^2\,d,
\label{eq:signature}
\end{equation}
where $n_\pm$ count the positive and negative eigenvalues and $(n_+,n_-)$ is the inertia. We call such a block fully reached. A Kramers doublet reads $(1,3)$, since all three components of its pseudospin are time odd. A doublet with $\Theta^2=1$ reads $(3,1)$, since it has one time-odd component and two time-even quadrupoles \cite{onoda2011,rau2019}. Every model with fully reached blocks thus has the same block dimensions and the same $\Theta^2$, so the inertia belongs to the process and not to the model. The hidden spin above is never reached, so its $\Theta^2$ never enters the kernel.

A spin $3/2$ with a twofold axis along $z$ reads $(1,1)$ [Fig.~\ref{fig:rule}(b)], and its zero signature leaves the sign undecided. The $\pi$ rotation about $z$ is then a symmetry of the emitter and the detector. Its eigenvalues $\pm i$ label the two states of each doublet and are exchanged by time reversal. This is the case of Wigner's classification in which Herring's character sum vanishes \cite{herring1937,zhang2015}. Every stretch then leaves a state diagonal in the basis of the rotation eigenstates, so the record reaches only the identity and the time-odd population difference of each doublet. In the three cases met here the signature divided by $d$ takes the values $1$, $-1$ and $0$ of the Frobenius--Schur indicator, so the record itself performs the test that separates real, quaternionic and complex structures \cite{baez2012}. The kernel is the record-statistics counterpart of the partial-inversion invariant that Pollmann and Turner compute from a quantum state \cite{pollmann2012}.

A classical model is microreversible if its labeled jumps obey detailed balance under a reversal of microstates, the classical form of Eq.~\eqref{eq:relation}. Each negative eigenvalue of the kernel then needs a pair of microstates exchanged by the reversal \cite{pra}. The tilted spin $3/2$ reads $(2,6)$, so a microreversible model needs at least twelve microstates while the quantum emitter has four levels. Quantum models of other stochastic processes show similar memory savings \cite{gu2012,thompson2018}.

Couple the bath to the electronic spin $3/2$ of an atom with nuclear spin $\tfrac12$, whose total spin is an integer. Without hyperfine coupling the record reaches only the electron, and the kernel shows Kramers blocks $(1,3)$. When the detector resolves the hyperfine levels, which are nondegenerate, every $\rt_x$ is proportional to a time-even projector and the kernel has no negative eigenvalue \cite{pra}.

\textit{Robustness.}---Losses, misreads and dark counts leave the inertia $(n_+,n_-)$ of the kernel unchanged as long as they treat each emission and its reversed absorption alike \cite{pra}. Missed jumps still act on the spin between the detected past and future. They change the eigenvalues of the kernel but cannot drive any of them through zero at an efficiency $\eta>0$ \cite{pra}. At $\eta=0.1$ the tilted spin $3/2$ still reads $(2,6)$ while its weakest eigenvalue falls from $1.1\times10^{-2}$ to $7.7\times10^{-5}$ of the largest, so losses cost only statistics. A relative mismatch $\delta$ between the efficiencies for emitted and absorbed quanta breaks reversal symmetry at order $\delta$ but adds spurious eigenvalues to the kernel only at order $\delta^2$ (End Matter~\ref{em:budget}). A mismatch thus shows up in the test of reversal symmetry while it is still too small to change the reading. In simulations the test flagged every mismatch large enough to change the reading. The largest mismatch that the reading survives is estimated at 0.1\% to 7\%, depending on the orientation of a spin $3/2$ and on the efficiency. A flagged mismatch is removed by discarding detected events of the more efficient kind at random.

\textit{Measurement protocol.}---The data are windows of length $2n$, a past $x$ followed by a future $\mirror y$, and the frequency of each window estimates $\mathcal K(x,y)$. The time reverse of the window $x\mirror y$ is $y\mirror x$, so comparing the two counts tests reversal symmetry and the symmetry of $\mathcal K$ at once. By Weyl's inequality each eigenvalue of the symmetrized estimate $\hat{\mathcal K}$ lies within $\|\hat{\mathcal K}-\mathcal K\|$ of its true value. Kernels estimated from two random halves of the shots differ by sampling noise alone, so their difference gauges this norm. Repeated splits set the threshold, and the eigenvalues above it in each block give the reading of Eq.~\eqref{eq:signature} (End Matter~\ref{em:budget}).

\begin{figure}[tbp]
\includegraphics[width=\columnwidth]{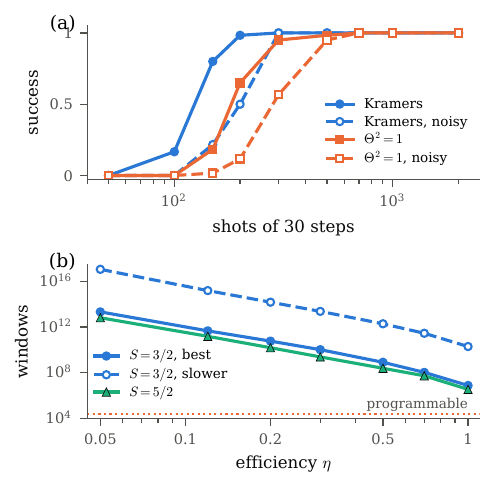}
\caption{(a) Probability that the protocol returns the full reading, $(1,3)$ for the programmed Kramers doublet and $(3,1)$ for its $\Theta^2=1$ control. Each point averages 60 simulated experiments, none of which returned the wrong class. Noisy runs add per step 3\% readout error, 1\% Pauli noise and 1\% amplitude damping of the data qubit. (b) Estimated number of independent detected windows needed for the full reading of an equilibrium spin against the detection efficiency $\eta$ of both emission and absorption. The two $S=3/2$ curves are the most favorable of 12 random orientations of the splitting axes and a slower one. The dotted line marks the programmed emitter of panel (a).}
\label{fig:budget}
\end{figure}

On a quantum processor a data qubit plays the Kramers pseudospin. A three-outcome ancilla is measured and reset at every step \cite{fossfeig2021,chertkov2022}, and each measurement records one outcome $q$. Odd steps apply a Kraus operator $u_q$ of a unital instrument, so $\rho_J=\openone/2$. Even steps apply $\sigma_yu_q^{\,T}\sigma_y=\Theta u_q^\dagger\Theta^{-1}$ with $\Theta=\sigma_yK$, so the record obeys Eq.~\eqref{eq:relation} with $\Theta^2=-1$. Absorptions and emissions between the two doublets of the thermal spin $3/2$ have this form \cite{pra}. The control replaces the even steps by $u_q^{\,T}=Ku_q^\dagger K$. It then has $\Theta=K$ with $\Theta^2=1$ and a kernel of the same rank, so only the sign of $\Theta^2$ separates it from the Kramers doublet. We choose the $u_q$ so that pasts and futures of two steps already reach the doublet fully. The weakest kernel eigenvalue is then $0.15$ of the largest, so few shots suffice. With 1000 shots of 30 steps each, every simulated experiment read $(1,3)$ for the Kramers doublet and $(3,1)$ for its control, with and without noise [Fig.~\ref{fig:budget}(a)].

A thermal spin $3/2$ or $5/2$ has much weaker kernel eigenvalues than the programmed emitter and needs of order $10^6$ to $10^{10}$ independent windows with a perfect detector [Fig.~\ref{fig:budget}(b)]. The detector must count emissions and absorptions with their angular momentum at zero field. Microwave photon counters detect single spins but work at finite field and miss the absorptions \cite{albertinale2021,wang2023}. Nondestructive detection of incoming photons could add the absorptions \cite{kono2018,niemietz2021}. Charge detectors count single electrons in both directions with equal efficiency but cannot see their spin \cite{fujisawa2006,kung2012}. Leads of opposite spin polarization supply the missing spin resolution if the detector tells the leads apart. A charge detector could thus read a Coulomb-blockaded dot at zero field through three pairs of such leads polarized along three non-coplanar axes. At lead polarization $0.4$ the full Kramers reading needs about $4\times10^5$ events \cite{pra}, so the remaining obstacles are technical.

\textit{Discussion.}---The selection rule and Kramers degeneracy are two consequences of one fact, the antisymmetry of $\langle\Theta\phi|\psi\rangle$ under exchange of $\phi$ and $\psi$ when $\Theta^2=-1$. The degeneracy applies it to eigenstates of $H$, while the rule applies it to the overlap of a predicted and a retrodicted state and so also needs equilibrium and a detector that resolves handedness. Weak antilocalization rests on the same antisymmetry as an interference correction averaged over disorder \cite{hikami1980,bergmann1984}. In a single classical record the antisymmetry instead forbids an event outright, much as photon antibunching from a single atom forbids a second click right after the first \cite{kimble1977}. In both cases the state left by the first event cannot produce the second.

The antisymmetry is the physical content of the nontrivial class of time reversal in $H^2(\mathbb Z_2^T,U(1))$. Other symmetry groups $G$ have classes in $H^2(G,U(1))$ and in higher cohomology, and whether records carry them is open. Even for time reversal the classical bound covers only microreversible models, and the memory that other classical models need is unknown. The hyperfine example suggests a direct test on two isotopes of one atom, one without nuclear spin and one with nuclear spin $\tfrac12$ and resolved hyperfine levels.

\textit{Data availability.}---The code that produces every number and figure in this Letter and in Ref.~\cite{pra} is openly available [TODO: repository or DOI].

\clearpage
\onecolumngrid
\begin{center}\textbf{End Matter}\end{center}
\twocolumngrid
\setcounter{secnumdepth}{1}
\appendix

\section{The selection rule and the signature}
\label{em:proofs}

\textit{Eq.~\eqref{eq:mirror}.} The reversed operations run each stretch backwards and form a channel \cite{crooks2008},
\begin{equation}
\tilde B_{m_1}\cdots\tilde B_{m_n}=\rho_J^{1/2}M_x^\dagger\rho_J^{-1/2},\;\; \sum_m\tilde B_m^\dagger\tilde B_m=\openone,
\end{equation}
for $x=m_1\cdots m_n$. This channel $\tilde{\mathcal E}(\rho)=\sum_m\tilde B_m\rho\tilde B_m^\dagger$ fixes $\rho_J$, since $\tilde{\mathcal E}(\rho_J)=\rho_J^{1/2}\big(\sum_mB_m^\dagger B_m\big)\rho_J^{1/2}=\rho_J$. Since $\tau$ is a bijection, Eq.~\eqref{eq:relation} gives $\tilde{\mathcal E}(\rho)=\Theta\,\mathcal E(\Theta^{-1}\rho\Theta)\,\Theta^{-1}$ with $\mathcal E(\rho)=\sum_mB_m\rho B_m^\dagger$, so $\tilde{\mathcal E}$ also fixes $\Theta\rho_J\Theta^{-1}$. When the stationary state is unique, as for the irreducible emitters considered here, $\Theta$ commutes with $\rho_J$. Equation~\eqref{eq:relation} gives $B_{\tau(m)}=e^{i\varphi_m}\Theta^{-1}\tilde B_m\Theta$. The time reverse $\mirror y$ of $y=m_1\cdots m_n$ applies $B_{\tau(m_n)}$ first and $B_{\tau(m_1)}$ last, so
\begin{equation}
M_{\mirror y}=e^{i\phi}\,\Theta^{-1}\rho_J^{1/2}M_y^\dagger\rho_J^{-1/2}\Theta
\end{equation}
with a global phase $\phi$. The effect of the future $\mirror y$ is then
\begin{equation}
E_{\mirror y}\equiv M_{\mirror y}^\dagger M_{\mirror y}=\Theta^{-1}\rho_J^{-1/2}\rho_y\,\rho_J^{-1/2}\,\Theta .
\label{eq:effect}
\end{equation}
With $\rho_J$ as prior it retrodicts $\rho_J^{1/2}E_{\mirror y}\rho_J^{1/2}=\Theta^{-1}\rho_y\Theta$, the time reverse of $\rho_y$. Since $\Theta$ commutes with $\rho_J$, the probability $P(x\,\mirror y)=\Tr[E_{\mirror y}\rho_x]$ is Eq.~\eqref{eq:mirror}.

If $x$ leaves a pure state $\psi$, then $\rt_x\propto|\chi\rangle\langle\chi|$ with $\chi=\rho_J^{-1/4}\psi$, and $P(x\,\mirror x)\propto|\langle\chi|\Theta\chi\rangle|^2$. Antiunitarity, $\langle\Theta a|\Theta b\rangle=\langle b|a\rangle$, gives
\begin{equation}
\langle\chi|\Theta\chi\rangle=\langle\Theta^2\chi|\Theta\chi\rangle=\Theta^2\langle\chi|\Theta\chi\rangle ,
\end{equation}
so $\langle\chi|\Theta\chi\rangle=0$ when $\Theta^2=-1$.

\textit{Eq.~\eqref{eq:signature}.} On Hermitian $d\times d$ matrices time reversal acts as
\begin{equation}
\mathcal T(Y)\equiv\Theta Y\Theta^{-1}=UY^{T}U^\dagger ,
\end{equation}
and $b(X,Y)=\Tr[X\mathcal T(Y)]$. The relation $UU^*=\Theta^2$ gives $U^T=\Theta^2U$, so $\mathcal T$ is self-adjoint for the Hilbert--Schmidt product and $b$ is real and symmetric. The map $\mathcal T$ is also an involution that preserves this product. Its eigenvalues $\pm1$ split each Hermitian $X$ into a time-even part $X_+$ and a time-odd part $X_-$, so $b(X,X)=\|X_+\|^2-\|X_-\|^2$. The form $b$ thus has inertia $(n_+,n_-)$ given by the dimensions of the two eigenspaces, and
\begin{equation}
n_+-n_-=\Tr\mathcal T=\sum_{ij}U_{ij}\overline{U_{ji}}=\Tr(UU^*)=\Theta^2d .
\end{equation}
The kernel is $\mathcal K=R^{T}bR$, with $b$ the matrix of the form in an orthonormal basis of Hermitian operators and $R$ the real-linear map from stretches to the operators $\rt_x$. By Sylvester's law its inertia is that of $b$ on the span of the $\rt_x$. On a fully reached block this span holds all $d^2$ Hermitian operators, so Eq.~\eqref{eq:signature} follows. On a partly reached block $b$ has at most as many positive and negative directions as on all Hermitian matrices. This gives $n_+\le d(d-1)/2$ for $\Theta^2=-1$ and $n_-\le d(d-1)/2$ for $\Theta^2=1$, so for a known $d=2$ two negative eigenvalues exclude $\Theta^2=1$. Blocks appear in the data as the connected components of the graph of pairs with $\mathcal K(x,y)>0$ \cite{pra}.

\textit{Canonical correlations.} With $D$ the diagonal matrix of the $P(x)$ on stretches that occur, and since $P(\mirror y)=P(y)$, the normalized kernel is
\begin{equation}
\big(D^{-1/2}\mathcal KD^{-1/2}\big)(x,y)=\frac{P(x\,\mirror y)}{\sqrt{P(x)P(\mirror y)}} .
\end{equation}
Let $1_x$ be the indicator of the event that the past reads $x$, and $1_{\mirror y}$ that of the future reading $\mirror y$. The entries of the normalized kernel are the second moments of $1_x/\sqrt{P(x)}$ and $1_{\mirror y}/\sqrt{P(\mirror y)}$. Its singular values are the canonical correlations between past and future. The matrix is symmetric, so they are the moduli of its eigenvalues. A normalized eigenvector $v$ with eigenvalue $\lambda$ defines a past variable $f=\sum_xv_x1_x/\sqrt{P(x)}$ and its time reverse $\mirror f=\sum_yv_y1_{\mirror y}/\sqrt{P(y)}$ in the future. Their expectations obey
\begin{equation}
\mathbb E[f\mirror f]=\lambda,\;\; \mathbb E[f^2]=\mathbb E[\mirror f^{\,2}]=1 .
\end{equation}
The variable $f$ is time even for $\lambda>0$ and time odd for $\lambda<0$. The eigenvector $v_x=\sqrt{P(x)}$ gives the constant variable $f=1$ with eigenvalue $1$, and all other variables have zero mean.

\section{The thermal emitter}
\label{em:emitter}

\textit{Exact instrument.} The jump operators $J_{q,\omega}=\sqrt{\gamma(\omega)}L_q(\omega)$ give the total jump rate
\begin{equation}
\Lambda\equiv\sum_{q,\omega}J_{q,\omega}^\dagger J_{q,\omega},
\end{equation}
which is block diagonal in the levels of $H$. Time reversal maps each $S_q$ to $-S_q^\dagger=(-1)^{q+1}S_{-q}$ and commutes with every $\Pi_\varepsilon$, so $\Lambda$ is invariant under $\Theta$. On a Kramers doublet $\{\psi,\Theta\psi\}$ a Hermitian $\Theta$-invariant operator $A$ is a multiple of the identity, since $\langle\Theta\psi|A|\Theta\psi\rangle=\langle\psi|A|\psi\rangle$ and $\langle\psi|A\Theta\psi\rangle=\langle\Theta^2\psi|\Theta A\psi\rangle=-\langle\psi|A\Theta\psi\rangle$. The levels of a generic $H$ are nondegenerate or Kramers doublets, so $\Lambda$ is scalar on every level. A jump of fixed energy creates no coherence between levels, and $\rho_J$ has none. The waiting-time integral then weights each level by the inverse of its total rate, and the outcomes are generated exactly by
\begin{equation}
B_{q,\omega}=J_{q,\omega}\Lambda^{-1/2},\;\; \rho_J\propto\Lambda e^{-\beta H}.
\end{equation}

\textit{Reversal relation.} For the Gibbs state $\rho_G\propto e^{-\beta H}$ detailed balance gives $\rho_G^{1/2}J_{q,\omega}^\dagger\rho_G^{-1/2}=\sqrt{\gamma(-\omega)}\,L_q(\omega)^\dagger$. Since $\rho_J\propto\Lambda\rho_G$ with $\Lambda$ scalar on each level,
\begin{equation}
\tilde B_{q,\omega}=\sqrt{\gamma(-\omega)}\,L_q(\omega)^\dagger\Lambda^{-1/2}=-\Theta B_{q,-\omega}\Theta^{-1}.
\end{equation}
The second equality uses $\Theta S_q^\dagger\Theta^{-1}=-S_q$, which follows from $\Theta\mathbf S\Theta^{-1}=-\mathbf S$ and antiunitarity, together with $[\Theta,H]=[\Theta,\Lambda]=0$. This is Eq.~\eqref{eq:relation} with $\tau(q,\omega)=(q,-\omega)$ and $U=e^{-i\pi S_y}$, so $\Theta^2=UU^*=(-1)^{2S}$ \cite{pra}.

\textit{Linear channels.} A detector of linear polarization couples through Hermitian operators $X$ with $\Theta X\Theta^{-1}=\pm X$. Between two Kramers doublets with bases $\{\psi_1,\Theta\psi_1\}$ and $\{\psi_2,\Theta\psi_2\}$ the block of $X$ obeys
\begin{equation}
X_{\rm b}=\begin{pmatrix}a&c\\\mp c^*&\pm a^*\end{pmatrix},\;\; X_{\rm b}X_{\rm b}^\dagger=\big(|a|^2+|c|^2\big)\openone ,
\end{equation}
with $a=\langle\psi_1|X|\psi_2\rangle$ and $c=\langle\psi_1|X|\Theta\psi_2\rangle$. Every Kraus operator acts on the pseudospin as a multiple of a unitary. The state $\rho_J$ is maximally mixed on each doublet, and a multiple of a unitary keeps it so. Each $\rho_x$ stays proportional to the projector of its doublet, so the $\rt_x$ span only the identity and each doublet reads $(1,0)$.

\section{Estimation and budgets}
\label{em:budget}

\textit{Threshold.} For a block estimated from $N$ independent windows, the variances of the entries in row $x$ of $\hat{\mathcal K}$ sum to about $P(x)/N$, so the sampling noise has spectral norm about $2\sqrt{\max_xP(x)/N}$. The estimator uses the symmetric part of $\hat{\mathcal K}$. The split-half difference $(\hat{\mathcal K}_A-\hat{\mathcal K}_B)/2$ carries the same noise and no signal. The threshold is $1.5$ times its largest spectral norm over six to ten random splits. With $d=2$ known, one eigenvalue of the wrong sign above threshold leaves the class undecided, while two of them read the wrong class. No simulation returned the wrong class.

\textit{Program.} The three Kraus operators $u_q$ of an odd step come from a random search over $2000$ unital three-outcome qubit instruments. A local refinement then improved whichever of the Kramers and control readings was harder. The $u_q$ are listed in the Supplemental Material of Ref.~\cite{pra}. Their weakest kernel eigenvalue is $0.15$ of the largest for both classes with stretches of two steps, whereas the median of $400$ random instruments needs about $270$ times more data \cite{pra}. With the absorption blocks of the thermal spin $3/2$ between its two doublets as the $u_q$, the full reading was reached in $7$ of $8$ simulations with $3\times10^4$ shots of $100$ steps and in $8$ of $8$ with $10^5$ shots.

\textit{Noise model.} Readout errors replace a label $q$ by one of the other two, equally for odd and even steps. After each step the data qubit suffers Pauli noise and amplitude damping, which we simulate as quantum trajectories. Pauli noise is unital and commutes with $\Theta$, so it commutes with the reversal and leaves the inertia unchanged \cite{pra}. Damping breaks the reversal weakly, and at $1\%$ per step it changed no reading [Fig.~\ref{fig:budget}(a)].

\textit{Thermal emitters.} The noise norm falls to half the weakest kernel eigenvalue $\lambda_{\rm weak}$ after
\begin{equation}
N^*=\Big(4\sqrt{\max_xP(x)}\big/|\lambda_{\rm weak}|\Big)^2
\label{eq:nstar}
\end{equation}
independent windows, which we evaluate from the exact spectrum. The simulated program counts the overlapping windows of shots of 30 steps and reaches a success probability of $0.8$ with as few as a third of its $N^*\approx1.1\times10^4$ windows, so the estimate is conservative. Over $12$ random orientations of a spin-$3/2$ splitting the median is $4.4\times10^8$ windows at $\eta=1$.

\textit{Efficiency mismatch.} Write $\mathcal K=A^TB$, where the columns of $A$ are the operators $\rho_x$ and those of $B$ the effects $E_{\mirror y}$ of Eq.~\eqref{eq:effect}. On a fully reached block the projector $Q$ onto the null space of the unperturbed kernel obeys $A_0Q=B_0Q=0$. A mismatch $\delta$ changes $A$ and $B$ at first order, $A=A_0+\delta A_1$ and $B=B_0+\delta B_1$, so
\begin{equation}
Q\mathcal KQ=\delta\,Q\big(A_1^TB_0+A_0^TB_1\big)Q+O(\delta^2)=O(\delta^2).
\end{equation}
The eigenvalues of the symmetric part $(\mathcal K+\mathcal K^T)/2$ that start at zero move only at order $\delta^2$ \cite{kato1995}. We take $\eta_{\rm abs}=(1-\delta)\eta_{\rm emit}$ with $\eta_{\rm emit}\in\{1,0.5,0.12\}$ for a spin $3/2$ in two orientations. The new eigenvalues and the asymmetry then scale as
\begin{equation}
\lambda_{\rm new}\approx c\,\delta^2\lambda_{\max},\;\; \|\mathcal K-\mathcal K^T\|\approx2\delta\lambda_{\max},
\end{equation}
with $\lambda_{\max}$ the largest eigenvalue and $c$ between $0.4$ and $0.5$. The new eigenvalues stay below $|\lambda_{\rm weak}|/4$ up to
\begin{equation}
\delta^*\approx\sqrt{|\lambda_{\rm weak}|/2\lambda_{\max}},
\end{equation}
which ranges from 0.1\% to 7\%. We sampled $10^7$ to $2\times10^{11}$ independent windows from the exact window distribution, with 20 runs per setting and $\delta\in\{0,0.1,0.3,0.5,1,3,10,20\}\%$. The reading stayed $(2,6)$ in at least 18 runs up to $\delta=3\%$ for the better orientation at $\eta_{\rm emit}=1$ and $0.5$, and up to $\delta=1\%$ in the other cases. At the next value on this grid it held in at most 6 runs. Every $\delta$ at which the reading failed exceeds the $\delta^*$ of its case. The test of reversal symmetry fires when the spectral norm of $(\hat{\mathcal K}-\hat{\mathcal K}^T)/2$ exceeds $1.5$ times that of its split-half difference. It never fired at $\delta=0$ and fired in every run with $\delta\ge1\%$, which includes every run whose reading changed.

\end{document}